\documentclass[a4paper,conference]{IEEEtran}
\IEEEoverridecommandlockouts
\usepackage[utf8]{inputenc}
\usepackage[T1]{fontenc}
\usepackage{microtype,newtxtext,newtxmath} 
\usepackage{mathtools,gensymb,eurosym,url,soul} 
\usepackage{algorithm}
\usepackage{algpseudocode}
\usepackage{caption}
\usepackage{subcaption}
\usepackage{longtable,booktabs,array}
\usepackage[backend=biber, sorting=none, style=ieee, maxnames=5]{biblatex}
\usepackage[switch]{lineno}
\usepackage{acronym}
\usepackage{graphicx}
\usepackage{xcolor}
\usepackage{xr} 
\usepackage{comment}
\usepackage{hyperref}
\hypersetup{colorlinks=true, breaklinks=true, 
  urlcolor=blue, linkcolor=blue,anchorcolor=blue,citecolor=blue,
  hypertexnames=true, final=true, 
  pdfpagemode = UseNone, 
  pdfauthor = {},
  pdftitle = {},
  pdfsubject = {},
  pdfkeywords = {}
}

\graphicspath{{img/} {imglocal/} {./}}
\DeclareGraphicsExtensions{.pdf,.png,.jpg,.mps}

\IEEEdisplaynontitleabstractindextext

\DeclareSourcemap{
  \maps[datatype=bibtex]{
    \map[overwrite=true]{
      \step[fieldset=url, null]
      \step[fieldset=doi, null]
      \step[fieldset=eprint, null]
    }
  }
}

\acrodef{ADC}[ADC]{Analog to Digital Converter}
\acrodef{ADEXP}[AdExp-I\&F]{Adaptive-Exponential Integrate and Fire}
\acrodef{AER}[AER]{Address-Event Representation}
\acrodef{AEX}[AEX]{AER EXtension board}
\acrodef{AE}[AE]{Address-Event}
\acrodef{AFM}[AFM]{Atomic Force Microscope}
\acrodef{AGC}[AGC]{Automatic Gain Control}
\acrodef{AI}[AI]{Artificial Intelligence}
\acrodef{ALD}[ALD]{Atomic Layer Deposition}
\acrodef{AMDA}[AMDA]{AER Motherboard with D/A converters}
\acrodef{ANN}[ANN]{Artificial Neural Network}
\acrodef{API}[API]{Application Programming Interface}
\acrodef{ARM}[ARM]{Advanced RISC Machine}
\acrodef{ASIC}[ASIC]{Application Specific Integrated Circuit}
\acrodef{AdExp}[AdExp-IF]{Adaptive Exponential Integrate-and-Fire}
\acrodef{BCI}[BCI]{Brain-Computer-Interface}
\acrodef{BCM}[BCM]{Bienenstock-Cooper-Munro}
\acrodef{BD}[BD]{Bundled Data}
\acrodef{BEOL}[BEOL]{Back-end of Line}
\acrodef{BG}[BG]{Bias Generator}
\acrodef{BMI}[BMI]{Brain-Machince Interface}
\acrodef{BTB}[BTB]{band-to-band tunnelling}
\acrodef{BP}[BP]{Back-propagation}
\acrodef{BPTT}[BPTT]{Back-propagation Through Time}
\acrodef{CAD}[CAD]{Computer Aided Design}
\acrodef{CAM}[CAM]{Content Addressable Memory}
\acrodef{CAVIAR}[CAVIAR]{Convolution AER Vision Architecture for Real-Time}
\acrodef{CA}[CA]{Cortical Automaton}
\acrodef{CCN}[CCN]{Cooperative and Competitive Network}
\acrodef{CDAC}[CDAC]{Capacitive Digital to Analog Converter}
\acrodef{CDR}[CDR]{Clock-Data Recovery}
\acrodef{CFC}[CFC]{Current to Frequency Converter}
\acrodef{CHP}[CHP]{Communicating Hardware Processes}
\acrodef{CMIM}[CMIM]{Metal-insulator-metal Capacitor}
\acrodef{CML}[CML]{Current Mode Logic}
\acrodef{CMP}[CMP]{Chemical Mechanical Polishing}
\acrodef{CMOL}[CMOL]{Hybrid CMOS nanoelectronic circuits}
\acrodef{CMOS}[CMOS]{Complementary Metal-Oxide-Semiconductor}
\acrodef{CNN}[CCN]{Convolutional Neural Network}
\acrodef{COTS}[COTS]{Commercial Off-The-Shelf}
\acrodef{CPG}[CPG]{Central Pattern Generator}
\acrodef{CPLD}[CPLD]{Complex Programmable Logic Device}
\acrodef{CPU}[CPU]{Central Processing Unit}
\acrodef{CSM}[CSM]{Cortical State Machine}
\acrodef{CSP}[CSP]{Constraint Satisfaction Problem}
\acrodef{CV}[CV]{Coefficient of Variation}
\acrodef{DAC}[DAC]{Digital to Analog Converter}
\acrodef{DAS}[DAS]{Dynamic Auditory Sensor}
\acrodef{DAVIS}[DAVIS]{Dynamic and Active Pixel Vision Sensor}
\acrodef{DBN}[DBN]{Deep Belief Network}
\acrodef{DBS}[DBS]{Deep-Brain Stimulation}
\acrodef{DFA}[DFA]{Deterministic Finite Automaton}
\acrodef{DIBL}[DIBL]{drain-induced-barrier-lowering}
\acrodef{DI}[DI]{delay insensitive}
\acrodef{DMA}[DMA]{Direct Memory Access}
\acrodef{DNF}[DNF]{Dynamic Neural Field}
\acrodef{DNN}[DNN]{Deep Neural Network}
\acrodef{DoF}[DoF]{Degrees of Freedom}
\acrodef{DPE}[DPE]{Dynamic Parameter Estimation}
\acrodef{DPI}[DPI]{Differential Pair Integrator}
\acrodef{DRRZ}[DR-RZ]{Dual-Rail Return-to-Zero}
\acrodef{DRAM}[DRAM]{Dynamic Random Access Memory}
\acrodef{DR}[DR]{Dual Rail}
\acrodef{DSP}[DSP]{Digital Signal Processor}
\acrodef{DVS}[DVS]{Dynamic Vision Sensor}
\acrodef{DYNAP}[DYNAP]{Dynamic Neuromorphic Asynchronous Processor}
\acrodef{EBL}[EBL]{Electron Beam Lithography}
\acrodef{ECoG}[ECoG]{Electrocorticography}
\acrodef{EDVAC}[EDVAC]{Electronic Discrete Variable Automatic Computer}
\acrodef{EEG}[EEG]{electroencephalography}
\acrodef{EIN}[EIN]{Excitatory-Inhibitory Network}
\acrodef{EM}[EM]{Expectation Maximization}
\acrodef{EPSC}[EPSC]{Excitatory Post-Synaptic Current}
\acrodef{EPSP}[EPSP]{Excitatory Post-Synaptic Potential}
\acrodef{ET}[ET]{Eligibility Trace}
\acrodef{EZ}[EZ]{Epileptogenic Zone}
\acrodef{FDSOI}[FDSOI]{Fully-Depleted Silicon on Insulator}
\acrodef{FEOL}[FEOL]{Front-end of Line}
\acrodef{FET}[FET]{Field-Effect Transistor}
\acrodef{FFT}[FFT]{Fast Fourier Transform}
\acrodef{FI}[F-I]{Frequency-Current}
\acrodef{FPGA}[FPGA]{Field Programmable Gate Array}
\acrodef{FR}[FR]{Fast Ripple}
\acrodef{FSA}[FSA]{Finite State Automaton}
\acrodef{FSM}[FSM]{Finite State Machine}
\acrodef{GIDL}[GIDL]{gate-induced-drain-leakage}
\acrodef{GOPS}[GOPS]{Giga-Operations per Second}
\acrodef{GPU}[GPU]{Graphical Processing Unit}
\acrodef{GUI}[GUI]{Graphical User Interface}
\acrodef{HAL}[HAL]{Hardware Abstraction Layer}
\acrodef{HFO}[HFO]{High Frequency Oscillation}
\acrodef{HH}[H\&H]{Hodgkin \& Huxley}
\acrodef{HMM}[HMM]{Hidden Markov Model}
\acrodef{HRS}[HRS]{High-Resistive State}
\acrodef{HR}[HR]{Human Readable}
\acrodef{HSE}[HSE]{Handshaking Expansion}
\acrodef{HW}[HW]{Hardware}
\acrodef{IBCI}[IBCI]{Implantable BCI}
\acrodef{ICT}[ICT]{Information and Communication Technology}
\acrodef{IC}[IC]{Integrated Circuit}
\acrodef{ICL}[ICL]{Implantable Closed Loop}
\acrodef{IDAC}[IDAC]{Current Digital to Analog Converter}
\acrodef{IEEG}[iEEG]{intracranial electroencephalography}
\acrodef{IF2DWTA}[IF2DWTA]{Integrate \& Fire 2--Dimensional WTA}
\acrodef{IFSLWTA}[IFSLWTA]{Integrate \& Fire Stop Learning WTA}
\acrodef{IF}[I\&F]{Integrate-and-Fire}
\acrodef{IMU}[IMU]{Inertial Measurement Unit}
\acrodef{INCF}[INCF]{International Neuroinformatics Coordinating Facility}
\acrodef{INI}[INI]{Institute of Neuroinformatics}
\acrodef{INRC}[Intel NRC]{Intel Neuromorphic Research Community}
\acrodef{IO}[I/O]{Input/Output}
\acrodef{IoT}[IoT]{Internet of Things}
\acrodef{IPSC}[IPSC]{Inhibitory Post-Synaptic Current}
\acrodef{IPSP}[IPSP]{Inhibitory Post-Synaptic Potential}
\acrodef{IP}[IP]{Intellectual Property}
\acrodef{ISI}[ISI]{Inter-Spike Interval}
\acrodef{IoT}[IoT]{Internet of Things}
\acrodef{JFLAP}[JFLAP]{Java - Formal Languages and Automata Package}
\acrodef{LEDR}[LEDR]{Level-Encoded Dual-Rail}
\acrodef{LFP}[LFP]{Local Field Potential}
\acrodef{LFSR}[LFSR]{Linear Feedback Shift Register}
\acrodef{LIF}[LIF]{Leaky Integrate and Fire}
\acrodef{LLC}[LLC]{Low Leakage Cell}
\acrodef{LNA}[LNA]{Low-Noise Amplifier}
\acrodef{LPF}[LPF]{Low Pass Filter}
\acrodef{LRS}[LRS]{Low-Resistive State}
\acrodef{LSM}[LSM]{Liquid State Machine}
\acrodef{LTD}[LTD]{Long Term Depression}
\acrodef{LTI}[LTI]{Linear Time-Invariant}
\acrodef{LTP}[LTP]{Long Term Potentiation}
\acrodef{LTU}[LTU]{Linear Threshold Unit}
\acrodef{LUT}[LUT]{Look-Up Table}
\acrodef{LVDS}[LVDS]{Low Voltage Differential Signaling}
\acrodef{MD}[MD]{Medical Device}
\acrodef{MCMC}[MCMC]{Markov-Chain Monte Carlo}
\acrodef{MEMS}[MEMS]{Micro Electro Mechanical System}
\acrodef{MFR}[MFR]{Mean Firing Rate}
\acrodef{MIM}[MIM]{Metal Insulator Metal}
\acrodef{ML}[ML]{Machine Leanring}
\acrodef{MLP}[MLP]{Multilayer Perceptron}
\acrodef{MOSCAP}[MOSCAP]{Metal Oxide Semiconductor Capacitor}
\acrodef{MOSFET}[MOSFET]{Metal Oxide Semiconductor Field-Effect Transistor}
\acrodef{MOS}[MOS]{Metal Oxide Semiconductor}
\acrodef{MRI}[MRI]{Magnetic Resonance Imaging}
\acrodef{NDFSM}[NDFSM]{Non-deterministic Finite State Machine} 
\acrodef{ND}[ND]{Noise-Driven}
\acrodef{NEF}[NEF]{Neural Engineering Framework}
\acrodef{NHML}[NHML]{Neuromorphic Hardware Mark-up Language}
\acrodef{NIL}[NIL]{Nano-Imprint Lithography}
\acrodef{NLP}[NLP]{Natural Language Processing}
\acrodef{NMDA}[NMDA]{N-Methyl-D-Aspartate}
\acrodef{NME}[NE]{Neuromorphic Engineering}
\acrodef{NN}[NN]{Neural Network}
\acrodef{NRZ}[NRZ]{Non-Return-to-Zero}
\acrodef{NSM}[NSM]{Neural State Machine}
\acrodef{OR}[OR]{Operating Room}
\acrodef{OTA}[OTA]{Operational Transconductance Amplifier}
\acrodef{PCB}[PCB]{Printed Circuit Board}
\acrodef{PCHB}[PCHB]{Pre-Charge Half-Buffer}
\acrodef{PCM}[PCM]{Phase Change Memory}
\acrodef{PD}[PD]{Parkinson Disease}
\acrodef{PE}[PE]{Phase Encoding}
\acrodef{PFA}[PFA]{Probabilistic Finite Automaton}
\acrodef{PFC}[PFC]{prefrontal cortex}
\acrodef{PFM}[PFM]{Pulse Frequency Modulation}
\acrodef{PM}[PM]{Personalized Medicine}
\acrodef{PR}[PR]{Production Rule}
\acrodef{PSC}[PSC]{Post-Synaptic Current}
\acrodef{PSP}[PSP]{Post-Synaptic Potential}
\acrodef{PSTH}[PSTH]{Peri-Stimulus Time Histogram}
\acrodef{PVD}[PVD]{Physical Vapor Deposition }
\acrodef{QDI}[QDI]{Quasi Delay Insensitive}
\acrodef{RAM}[RAM]{Random Access Memory}
\acrodef{RDF}[RDF]{random dopant fluctuation}
\acrodef{RELU}[ReLu]{Rectified Linear Unit}
\acrodef{RLS}[RLS]{Recursive Least-Squares}
\acrodef{RMSE}[RMSE]{Root Mean Squared-Error}
\acrodef{RMS}[RMS]{Root Mean Squared}
\acrodef{RNN}[RNN]{Recurrent Neural Network}
\acrodef{ROLLS}[ROLLS]{Reconfigurable On-Line Learning Spiking}
\acrodef{RRAM}[R-RAM]{Resistive Random Access Memory}
\acrodef{R}[R]{Ripples}
\acrodef{SAC}[SAC]{Selective Attention Chip}
\acrodef{SAT}[SAT]{Boolean Satisfiability Problem}
\acrodef{SCX}[SCX]{Silicon CorteX}
\acrodef{SD}[SD]{Signal-Driven}
\acrodef{SDSP}[SDSP]{Spike Driven Synaptic Plasticity}
\acrodef{SEM}[SEM]{Spike-based Expectation Maximization}
\acrodef{SLAM}[SLAM]{Simultaneous Localization and Mapping}
\acrodef{SNN}[SNN]{Spiking Neural Network}
\acrodef{SNR}[SNR]{Signal to Noise Ratio}
\acrodef{SOC}[SOC]{System-On-Chip}
\acrodef{SOI}[SOI]{Silicon on Insulator}
\acrodef{SoA}[SoA]{state-of-the-art}
\acrodef{SP}[SP]{Separation Property}
\acrodef{SRAM}[SRAM]{Static Random Access Memory}
\acrodef{STDP}[STDP]{Spike-Timing Dependent Plasticity}
\acrodef{STD}[STD]{Short-Term Depression}
\acrodef{STEM}[STEM]{Science, Technology, Engineering and Mathematics}
\acrodef{STP}[STP]{Short-Term Plasticity}
\acrodef{STT-MRAM}[STT-MRAM]{Spin-Transfer Torque Magnetic Random Access Memory}
\acrodef{STT}[STT]{Spin-Transfer Torque}
\acrodef{SW}[SW]{Software}
\acrodef{TCAM}[TCAM]{Ternary Content-Addressable Memory}
\acrodef{TFT}[TFT]{Thin Film Transistor}
\acrodef{TPU}[TPU]{Tensor Processing Unit}
\acrodef{TRL}[TRL]{Technology Readiness Level}
\acrodef{USB}[USB]{Universal Serial Bus}
\acrodef{VHDL}[VHDL]{VHSIC Hardware Description Language}
\acrodef{VLSI}[VLSI]{Very Large Scale Integration}
\acrodef{VOR}[VOR]{Vestibulo-Ocular Reflex}
\acrodef{WCST}[WCST]{Wisconsin Card Sorting Test}
\acrodef{WTA}[WTA]{Winner-Take-All}
\acrodef{XML}[XML]{eXtensible Mark-up Language}
\acrodef{CTXCTL}[CTXCTL]{CortexControl}
\acrodef{divmod3}[DIVMOD3]{divisibility of a number by three}
\acrodef{hWTA}[hWTA]{hard Winner-Take-All}
\acrodef{sWTA}[sWTA]{soft Winner-Take-All}
\acrodef{APMOM}[APMOM]{Alternate Polarity Metal On Metal}
\usepackage{arydshln}
\usepackage{nicematrix}

\begin{document}

\title{SPAIC: A sub-$\mu$W/Channel, 16-Channel
General-Purpose Event-Based Analog Front-End
with Dual-Mode Encoders
\thanks{This work received funding from the European Union's H2020 research and innovation programme under the H2020 BeFerrosynaptic (871737), MeM-Scales (871371) projects.}
}
\author{\IEEEauthorblockN{
Shyam Narayanan\IEEEauthorrefmark{1},
Matteo Cartiglia\IEEEauthorrefmark{1}
Arianna Rubino\IEEEauthorrefmark{1},
Charles Lego \IEEEauthorrefmark{1},
Charlotte Frenkel\IEEEauthorrefmark{2},
Giacomo Indiveri\IEEEauthorrefmark{1}}
\IEEEauthorblockA{
\IEEEauthorrefmark{1}Institute of Neuroinformatics,
University of Zurich and ETH Zurich, Switzerland  \\
\IEEEauthorrefmark{2} TU Delft, Netherlands
\\
Email: \{shyam, camatteo, giacomo\}@ini.uzh.ch}
}

\newcommand{\MC}[1]{{\color{red}  Matteo: #1}}
\newcommand{\GI}[1]{{\color{blue}  Giacomo: #1}}
\newcommand{\SN}[1]{{\color{orange}  Shyam: #1}}
\newcommand{\CF}[1]{{\color{forestgreen}  Charlotte: #1}}
\newcommand{\AR}[1]{{\color{teal}  Arianna: #1}}

\maketitle

\begin{abstract}
  Low-power event-based analog front-ends (AFE) are a crucial component required to build efficient end-to-end neuromorphic processing systems for edge computing.
  Although several neuromorphic chips have been developed for implementing spiking neural networks (SNNs) and solving a wide range of sensory processing tasks, there are only a few general-purpose analog front-end devices that can be used to convert analog sensory signals into spikes and interfaced to neuromorphic processors.
  In this work, we present a novel, highly configurable analog front-end chip, denoted as “SPAIC” (signal-to-spike converter for analog AI computation), that offers a general-purpose dual-mode analog signal-to-spike encoding with delta modulation and pulse frequency modulation, with tunable frequency bands.
  The ASIC is designed in a 180\,nm process.
  It supports and encodes a wide variety of signals spanning 4 orders of magnitude in frequency, and provides an event-based output that is compatible with existing neuromorphic processors.
  We validated the ASIC for its functions and present initial silicon measurement results characterizing the basic building blocks of the chip.
\end{abstract}


\section{Introduction}
Spiking Neural Networks (SNNs) represent a powerful low-power event-based processing computing paradigm for processing streaming data on the edge \cite{Bartolozzi_etal22}.

\begin{figure}
  \centering
  \includegraphics[width=0.4\textwidth]{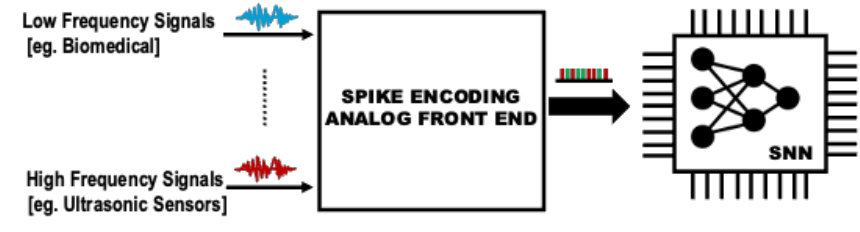}
  \caption{Illustration of sensory signal processing pipeline using a general purpose analog front end and SNN.}
  \label{fig:illust}
\end{figure}

To best exploit this novel emerging computing paradigm and build a robust end-to-end SNN sensory processing pipeline, designing efficient event-based analog front-ends is paramount. Fig. \ref{fig:illust} illustrates such an end-to-end pipeline.
There are various methods of encoding analog signals to spikes~\cite{WenzheEncodingFNS21, EvelinaEncodingFNS22}. On the hardware front, several analog-to-spike encoders have been developed and demonstrated on silicon, either using Delta Modulation (DM) schemes~\cite{YangISSCC16}, thereby encoding the temporal changes in the original signal with spike timing, or with  Pulse-Frequency Modulation (PFM) schemes~\cite{Corradi_Indiveri15, YangISSCC19, Badami15}, thereby encoding the amplitude of the signal with spike rates.
However, these encoders were always optimized for a specific  application and for the corresponding frequency bands.

\begin{figure}[b]
  \centering
  \includegraphics[width=0.38\textwidth]{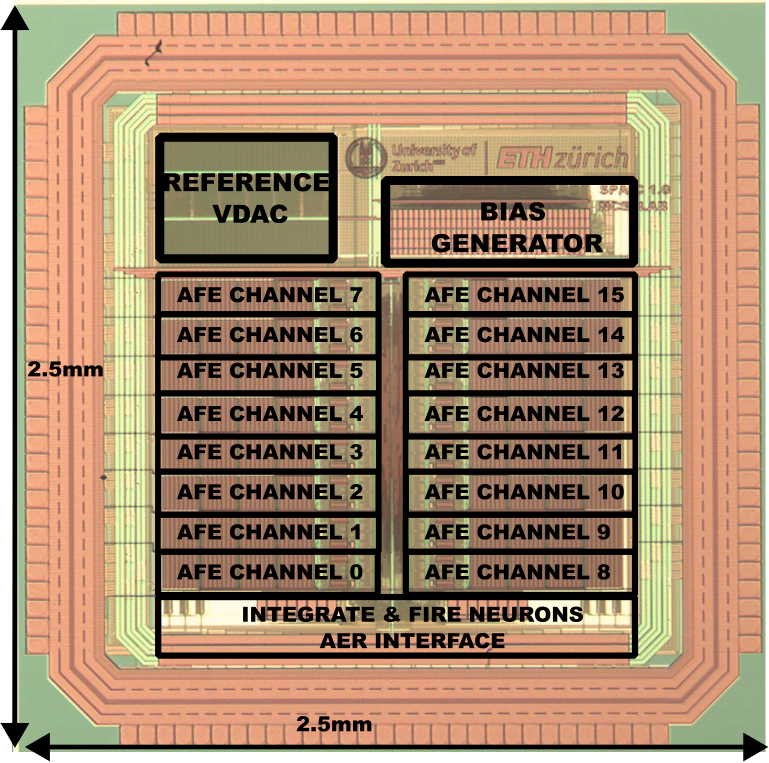}
  \caption{SPAIC chip micrograph.}
  \label{fig:spaiclay}
\end{figure}

To the best of our knowledge, no general-purpose solution has been proposed to allow exploration and prototyping with existing SNN neuromorphic computing platforms.
In this work, we present a highly configurable analog front-end ASIC, denoted as ``SPAIC'' (signal to spike converter for analog AI computation), which is compatible with existing neuromorphic processors~\cite{MoradiDynapse, DaviesLoihi, Frenkel_Indiveri22, Mayr_etal19, Frenkel_etal19}, and which offers a general-purpose dual-mode (DM and PFM) analog-to-spike encoding with tunable frequency bands (see Fig.~\ref{fig:spaiclay} for the chip micrograph).

\section{ASIC architecture}
\label{sec:arch}

\begin{figure*}
  \centering
  \includegraphics[width=0.98\textwidth]{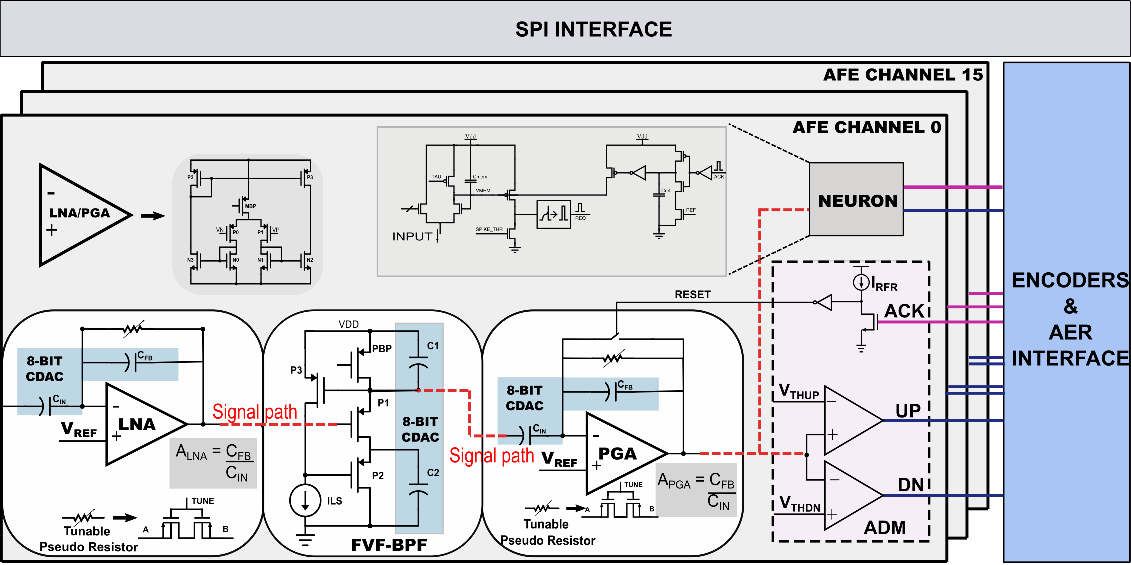}
  \caption{Architecture of SPAIC Analog Front-End ASIC.}
  \label{fig:arch}
\end{figure*}

SPAIC comprises 16 identical analog channels feeding a common Address Event Representation (AER) interface as shown in Fig.~\ref{fig:arch}. Each channel has four stages: a low noise amplifier section (LNA), a fourth-order flipped voltage follower (FVF) bandpass filter, a programmable gain stage (PGA), and the encoding stage. The LNA amplifies weak signals with tunable gain of 0 to 24dB. The FVF BPF filters the amplified signal with a tunable center frequency and Q~\cite{XuFVFFilter}. The PGA further amplifies the filtered signal with a tunable gain up to 24dB. The amplified signal is then split in two mutually exclusive paths where either the asynchronous delta modulator encodes the temporal derivative of the amplified signal, or the integrate and fire neuron~\cite{AriannaNeuron} is enabled and the signal is encoded as a pulse-frequency-modulated spike train. All configurations to the chip are provided in three ways: a configurable on-chip bias generator that generates the necessary bias currents for all circuits in the analog front-end, an 8-bit capacitor DAC (CDAC) to tune the filter parameters and a voltage DAC (VDAC) to set the UP and DN thresholds for the Asynchronous Delta Modulator (ADM). All three of these DACs are configurable via Serial Peripheral Interface (SPI) protocol. 

\section{Circuit implementation}
The SPAIC ASIC was designed and fabricated in a bulk 180\,nm technology node. The chip dimensions including the seal ring are 2.5\,mm$\times$2.5\,mm. The circuit implementation of the major building blocks is described hereafter.

\subsection{Low noise amplifiers}
For modularity reasons, both amplifiers in the analog front-end (LNA and PGA) are built on the same Operational Tranconductance Amplifier (OTA) core with minor changes adapted for noise and power.
The structure of the OTA is based on a well-known wide input range current-mirror-type transconductance amplifier~\cite{HarrisonOTA}.
 The primary reason for this design choice was to accommodate for large input changes that may or may not be present depending on the input sensor. The amplifiers are operated in a closed loop as capacitive feedback amplifiers with a DC-Servo loop (DSL) implemented with tunable pseudo resistors as shown in Fig.~\ref{fig:arch}. The capacitance ratio of the feedback to input capacitance determines the overall closed-loop gain. This gain configuration is implemented as a 4-bit binary weighted capacitance DAC, leading to 16 distinct gain values. The gains as well as the noise performance were validated on silicon and are described in Section \ref{section:meas}.
\subsection{Flipped Voltage Follower-based bandpass filter}
Every AFE channel has a 4th-order BPF with tunable center frequency and Q. The FVF-based filter topology shown in Fig.~\ref{fig:arch} helps achieve better noise performance due to the inherent current reuse present in the architecture.

The filter’s center frequency ($\omega_{0}$) is tunable with the on-chip bias generator. The capacitors of this filter are implemented as an 8-bit CDAC. By configuring $C_1$ and $C_2$, the Q and the center frequency can be adjusted in a precise manner with a resolution of $C_{1,2}/256$ steps, as described in Eq.~(\ref{eq:2}).
\begin{equation}
    \omega_0 = \sqrt \frac { gm_1 \cdot gm_2}{C_1\cdot C_2}, \\
    Q = \sqrt \frac { gm_2 \cdot C_2}{gm_1\cdot C_1}
    \label{eq:2}
\end{equation}
As each filter’s center frequency and Q are programmable, they can be configured as a parallel filter bank or as identical parallel electrode interface channels.
The frequency response, when configured as a filter bank, measured from the functional silicon is shown in Fig.~\ref{fig:filterbank}, Section~\ref{section:meas}. 

\subsection{Asynchronous Delta Modulation encoder}
The ADM is designed based on the foundation of a level-crossing ADC. The amplified and conditioned signal is compared with two known voltage thresholds (an up threshold and a down threshold) set by an on-chip voltage DAC. When the signal crosses the up threshold, an "UP" spike is generated and similarly, in the other direction a "DOWN" spike is generated when the signal goes below the set down threshold. The DACs were simulated running Monte Carlo analysis and the absolute accuracy (3$\sigma$) of the voltage thresholds was found to be $\approx \pm$ 900 $\mu V$. The comparator used in this ADM was designed with hysteresis~\cite{AllstotComparator,allen2011cmos} to avoid false triggering due to fluctuations in the signal path arriving at the input of the comparator. 

\subsection{Pulse Frequency Modulation encoder}
The PFM encoder was designed using a leaky integrate and fire neuron (LIF) circuit~\cite{AriannaNeuron} which inherently encodes the amplitude of the input current into pulse frequency. Therefore, before being encoded into spikes, the input signal is half-wave rectified and converted into current by a wide-range transconductance amplifier. The output current of this amplifier flows into the input of the neuron which generates a spike when the input current crosses an externally set spiking threshold of the neuron. 


%
\section{Silicon validation and measurement results} \label{section:meas}

\begin{figure}
  \centering
  \includegraphics[width=0.45\textwidth]{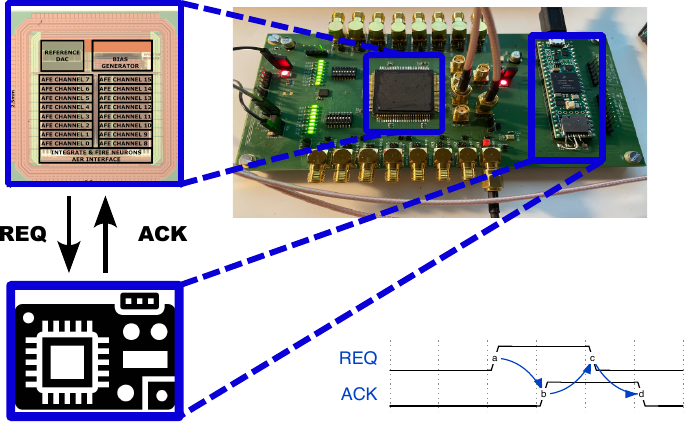}
  \caption{Measurement setup of the ASIC with a Teensy microcontroller, communicating via a 4-phase asynchronous handshaking protocol.}
  \label{fig:teensy}
\end{figure}

A custom evaluation board with a Teensy microcontroller was designed to evaluate the functionality of the ASIC. The setup is shown in Fig.~\ref{fig:teensy}. The microcontroller communicates with the ASIC and programs its internal registers via an SPI communication protocol. 
The two separate encoding paths are fully independent of each other and have their own AER circuitry. The AER circuitry operates based on a well-established four-phase hand-shaking mechanism described in Fig.~\ref{fig:teensy} \cite{Lazzaro_etal93}. The asynchronous nature of the AER interface ensures sparse event-driven communication with other asynchronous devices. 

\begin{figure}
  \centering
  \includegraphics[width=0.45\textwidth]{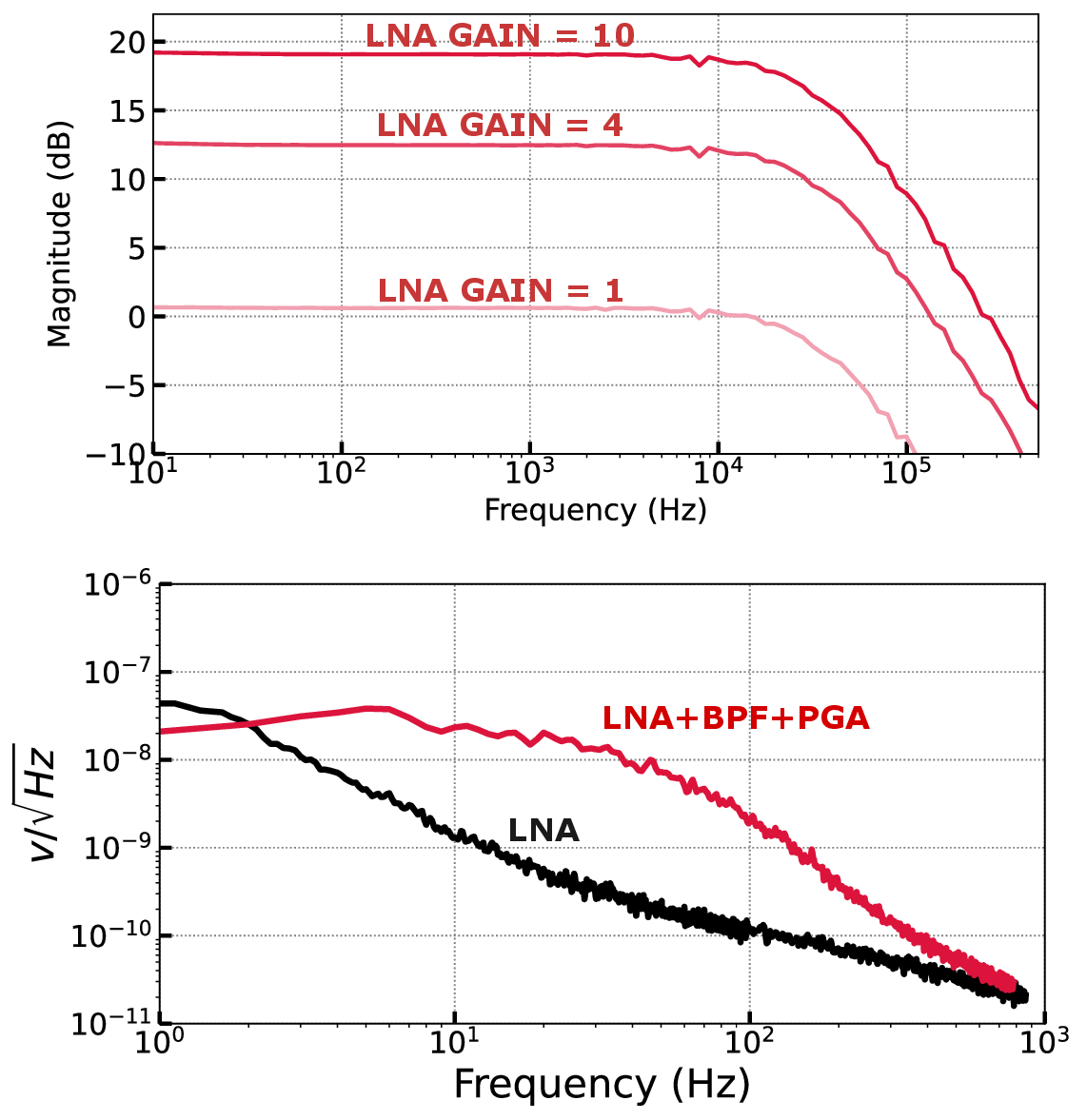}
  \caption{Frequency Response of LNA (top)
noise power spectral density (PSD) measurement (bottom).}
  \label{fig:npsd}
\end{figure}

\begin{figure}
  \centering
  \includegraphics[width=0.45
\textwidth]{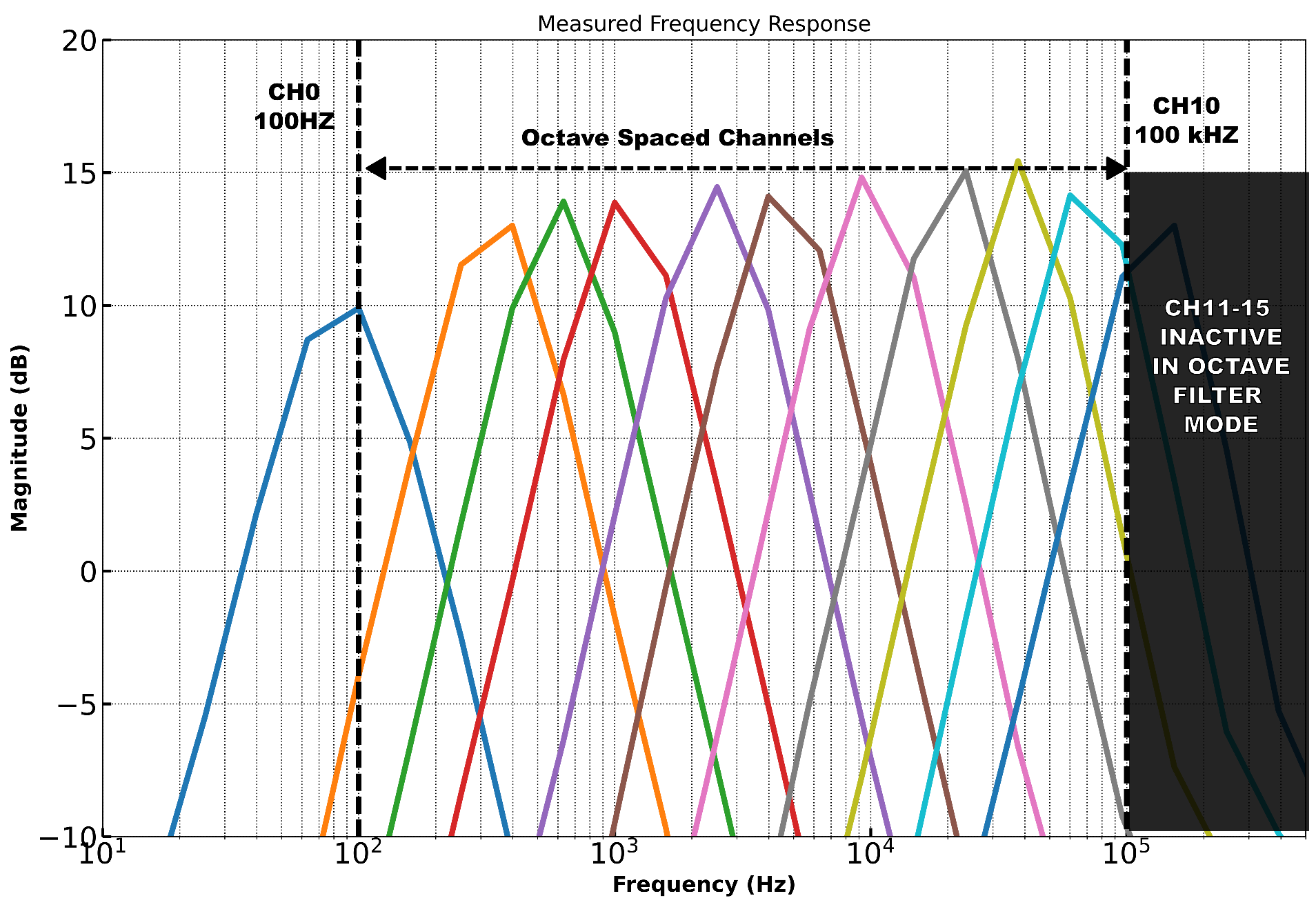}
  \caption{Frequency Response of the bandpass filter bank}
  \label{fig:filterbank}
\end{figure}

\begin{figure}
  \centering
  \includegraphics[width=0.45\textwidth]{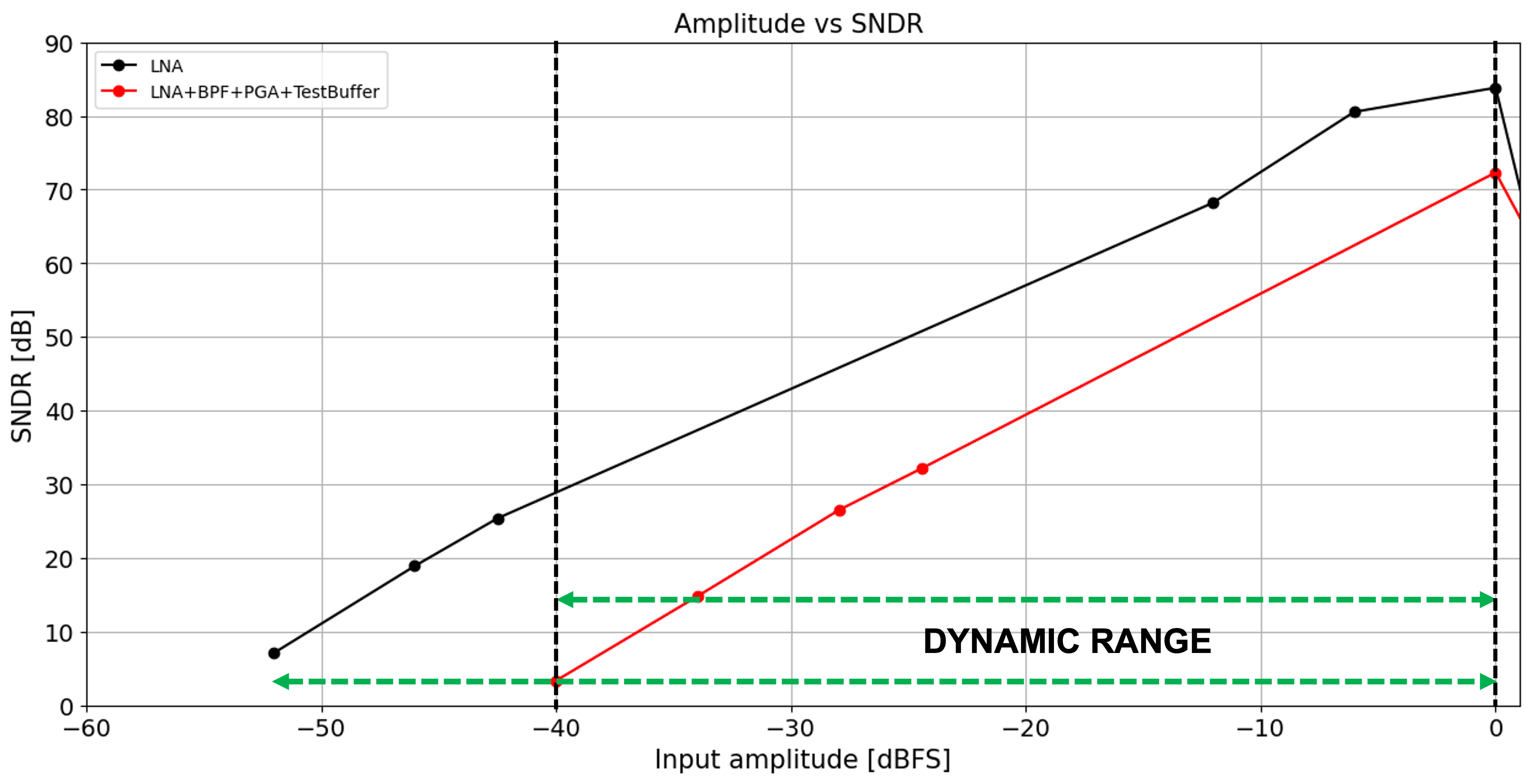}
  \caption{SNDR Measurement of an AFE Channel.}
  \label{fig:sndr}
\end{figure}

\begin{figure}
  \centering
  \includegraphics[width=0.45\textwidth]{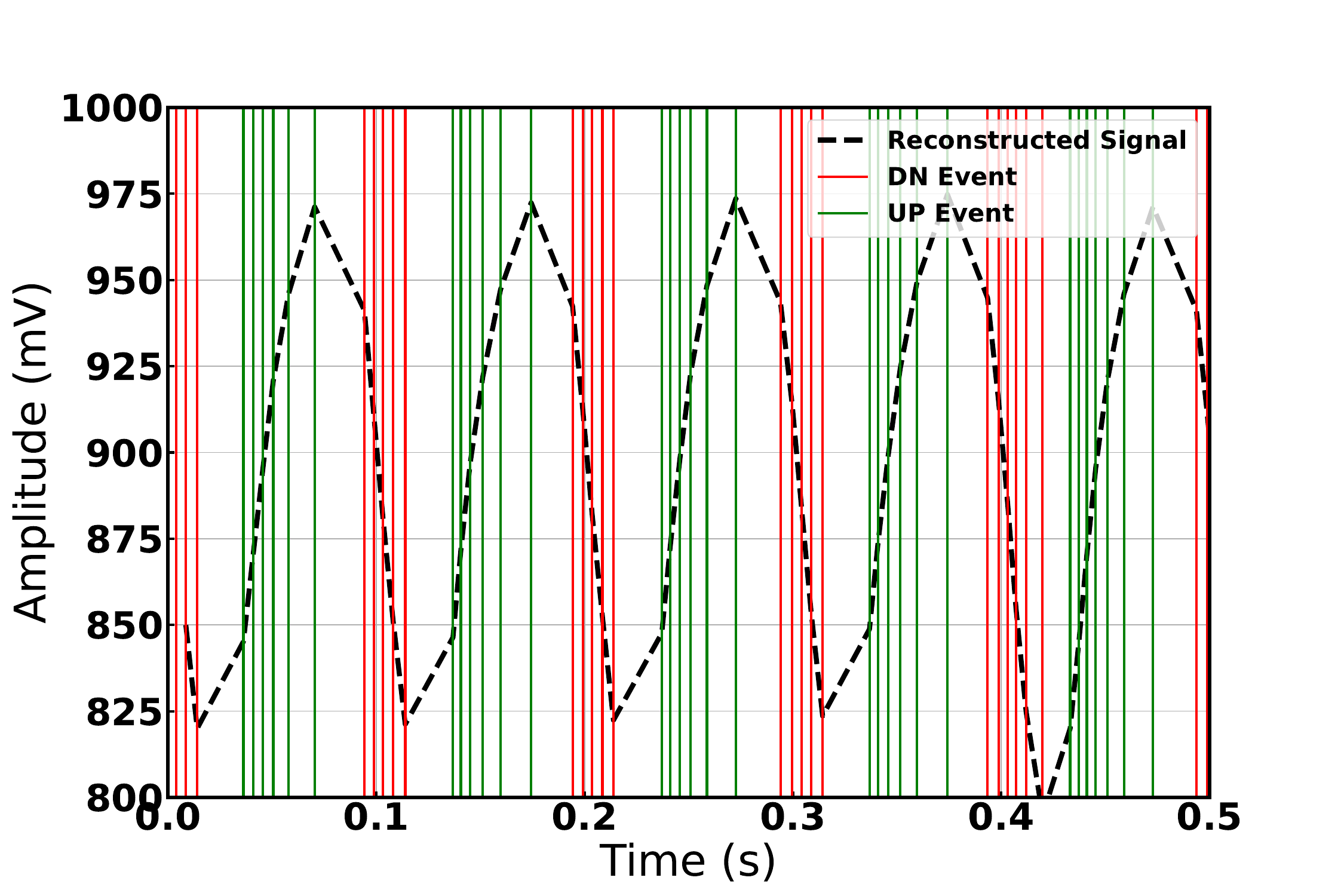}
  \caption{ADM spiking response to a 100Hz signal and reconstruction of the signal from spikes.}
  \label{fig:admresp}
\end{figure}

\begin{figure}
  \centering
  \includegraphics[width=0.45\textwidth]{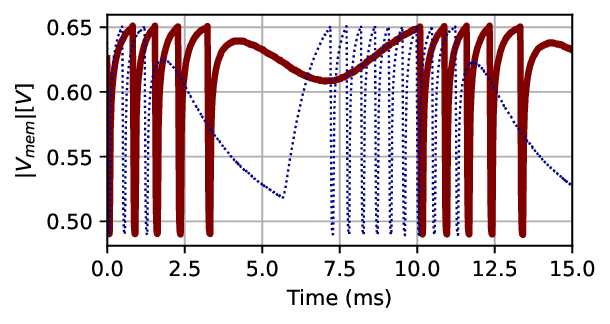}
  \caption{Neuron's membrane potential response for different amplitudes of the input signal of the AFE. Red (100mV), Blue (400mV).}
  \label{fig:membrane}
\end{figure}

\begin{figure}
  \centering
  \includegraphics[width=0.45\textwidth]{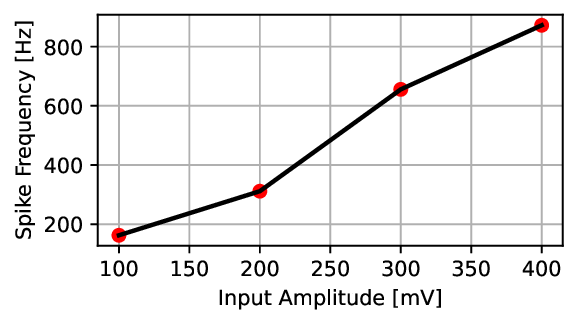}
  \caption{Spike frequency response to different input amplitudes.}
  \label{fig:spkresp}
\end{figure}

 The frequency response of the amplifiers and the noise power spectral density (Noise-PSD) measurement of the AFE are shown in Fig.~\ref{fig:npsd}. The plots show the noise PSD in the 1\,Hz-1\,kHz band, which is mainly dominated by flicker noise. The input-referred noise is 1.4\,$\mu V$ measured at the output of the first LNA.
Fig~\ref{fig:filterbank} shows a measurement where the bias currents were programmed in octaves to obtain center frequencies spaced in octaves.
 When the filter channels are octave-spaced, 11 channels can cover a range of 100\,Hz-100\,kHz. When used as identical channels, all 16 curves roughly overlap each other. The filter has a dynamic range superior to 40dB, however, reducing the gain of the first amplification stage helps to avoid distortion in the filtering stage. The silicon measurement of signal-to-noise and distortion ratio (SNDR) of a single AFE channel reveals a dynamic range of 42dB at the output of PGA as shown in Fig.~\ref{fig:sndr}.
The encoding functionality of both ADM and PFM were measured on the silicon. A 100\,Hz pure tone was encoded with ADM and the UP and DOWN events were recorded. To validate the encoding, the signal was reconstructed from the spikes and is shown in Fig.~\ref{fig:admresp}. A finer or coarser signal reconstruction can be achieved depending on the configured voltage thresholds. The final encoding accuracy is further determined by the input referred offset of the comparators. 
Similarly, the functionality of the neuron as a PFM encoder was validated by physical measurement of its membrane potential in response to a 100Hz pure tone as shown in Fig.~\ref{fig:membrane}. The red trace in Fig.~\ref{fig:membrane} shows the response to a 100mV input and blue trace to a 400mV input. The almost linear relationship of the spiking frequency to input amplitude shown in Fig.~\ref{fig:spkresp} confirms the intended functionality of the block. 

Table~\ref{tab::specs} compares the performance of this work with similar event-based analog front-ends. This design can cover a programmable range from 100\,Hz to 100\,KHz.
\begin{table}[htbp!]
\caption{Comparision with prior work}
\centering
\resizebox{\linewidth}{!}{%
\begin{NiceTabular}{ ccccc }

    		\textbf{}
            &\textbf{This Work}
    		&\textbf{\cite{YangISSCC19}}
    		&\textbf{\cite{YangISSCC16}}
            &\textbf{\cite{Badami15}}\\
		  \hline
            
            Technology [nm]  & 180 & 180 & 180 & 90\\
            \hdottedline
            
            Feature & Analog to &  Analog to & Analog to & Analog to \\& Events & Events & Events & Events \\
             
            \hdottedline
            
            Channels &  16 & 16 & 64x2 & 16\\
         
        \hdottedline
        
		  Bandwidth [Hz] &  $<$100-100K & 100-5K & 8-20K & 75-5K\\
        
       \hdottedline
       
		  Power/Channel [nW] &  $<$800@100KHz & 24 & 430 & 380\\
        
        \hdottedline

        Normalized Power \footnotemark[1][nW]  &  532 & 88 & 821 & 1500\\
        
        \hdottedline
		  Input Referred Noise [$\mu V_{rms}$] &  1.4@LNA & 8.3 & 30 & 32.5\\
         
        \hdottedline
        
		  Dynamic Range [dB] &  ~55@LNA & ~40@IAF & ~55@BPF & ~45@LNA \\ &~40@PGA\\
        
         \hdottedline
         
		  Area/Channel [$mm^{2}$] &  0.09 & 0.1 & 0.26 & 0.13\\
         
        \hdottedline
        
		  Functional Building &  LNA & LNA & LNA & LNA\\
            Blocks & BPF & BPF & BPF & BPF \\
                   & PGA & FWR & PGA & FWR \\
                   & ADM & IAF & ADM & LPF \\
                   & OTA+AdExpI\&F & & & \\
                    \hline
\end{NiceTabular}}
\footnotemark{Normalized to 5kHz using equation (8) in reference \cite{YangISSCC16}}\\
\label{tab::specs}
  \vspace*{-0.2cm}

\end{table}
The AFE consumes about 800\,nW with the channel processing at the highest frequency (100\,kHz) enabled. All other lower-frequency channels consume proportionally lower power. The power consumption was normalized to 5kHz and the normalized power for SPAIC AFE is about 532 nW.

\section{Conclusions}
In this paper, we proposed a highly configurable general-purpose signal-to-spike encoding ASIC with a dual-mode encoding scheme, designed and fabricated in 180\,nm technology. 

We demonstrated a generic signal conditioning and dual-way encoding scheme that one can pair with a spiking neural network to build an event-based neuromorphic sensing system. 
The design is comparable to the state-of-the-art in terms of dynamic range and silicon area and is better in terms of the operating range across a larger frequency range and noise performance. 

Future work includes quantitatively benchmarking the encoding schemes with a few selected applications such as low-frequency biomedical signals to auditory signals and high-frequency ultrasonic signals and quantifying the normalized power consumption for any selected application.

This work represents a key enabler for building an end-to-end signal acquisition pipeline for edge computing nodes with spiking neural networks. 
\newpage
\printbibliography 
\end{document}